\documentclass{article}

\usepackage{arxiv}
\usepackage[utf8]{inputenc} 
\usepackage[T1]{fontenc}    
\usepackage{hyperref}       
\usepackage{url}            
\usepackage{booktabs}       
\usepackage{amsfonts}       
\usepackage{nicefrac}       
\usepackage{microtype}      
\usepackage{graphicx}
\usepackage{doi}
\usepackage{cite}
\usepackage{amsmath,amssymb,amsfonts}
\usepackage{algorithmic}
\usepackage{graphicx}
\usepackage{textcomp}
\usepackage{xcolor}
\usepackage{placeins}
\usepackage{algorithm}
\usepackage{algorithmic}

\usepackage{amsthm}
\newtheorem{lemma}{Lemma}
\newtheorem{definition}{Definition}

\newtheorem{remark}{Remark}
\newtheorem{proposition}{Proposition}

\newtheorem{corollary}{Corollary}

\def\BibTeX{{\rm B\kern-.05em{\sc i\kern-.025em b}\kern-.08em
    T\kern-.1667em\lower.7ex\hbox{E}\kern-.125emX}}

\title{A Novel Approach to Differential Privacy with Alpha Divergence}
\author{
Yifeng Liu\\
\textit{Department of Electrical and Computer Engineering} \\
\textit{The University of British Columbia, Vancouver}\\
Vancouver, Canada \\
lyf666@student.ubc.ca
\and
Zehua Wang\textsuperscript{*}\\
\thanks{Corresponding author, This work was supported part by the Natural Sciences and Engineering Research Council (NSERC) of Canada under Discovery Grants RGPIN2021-02970 and DGECR-2021-00187.}
\textit{Department of Electrical and Computer Engineering}\\
\textit{The University of British Columbia, Vancouver}\\
Vancouver, Canada \\
zwang@ece.ubc.ca}

\begin{document}
\maketitle
\begin{abstract}
As data-driven technologies advance swiftly, maintaining strong privacy measures becomes progressively difficult. Conventional $(\epsilon, \delta)$-differential privacy, while prevalent, exhibits limited adaptability for many applications. To mitigate these constraints, we present alpha differential privacy (ADP), an innovative privacy framework grounded in alpha divergence, which provides a more flexible assessment of privacy consumption. This study delineates the theoretical underpinnings of ADP and contrasts its performance with competing privacy frameworks across many scenarios. Empirical assessments demonstrate that ADP offers enhanced privacy guarantees in small to moderate iteration contexts, particularly where severe privacy requirements are necessary. The suggested method markedly improves privacy-preserving methods, providing a flexible solution for contemporary data analysis issues in a data-centric environment.
\end{abstract}

\section{Introduction}
In the modern data-centric age, protecting individual privacy has emerged as a critical issue for researchers, practitioners, and legislators. Conventional data security techniques frequently fail to maintain an optimal equilibrium between data utility and personal privacy. Differential privacy, initiated by Dwork et al. \cite{Dwork2006}, has gained importance as a standard for privacy-preserving data analysis, supported by strong theoretical guarantees.

The fundamental premise of differential privacy is indistinguishability, which guarantees that the results of studies on datasets differing by one individual are statistically indistinguishable \cite{Dwork2006_icalp}. This essential attribute has enabled the extensive implementation of differential privacy in multiple fields, such as machine learning, statistical analysis, and data mining \cite{Abadi2016}, \cite{wang2019collecting}, \cite{friedman2010data}. Despite various follow-up studies \cite{Balle2018}, \cite{Kairouz2015} focused on enhancing its efficacy, the traditional $(\epsilon, \delta)$-differential privacy framework may still be insufficiently adaptable to meet more nuanced privacy demands or to effectively reconcile the trade-offs between privacy and utility.

Researchers have been actively seeking privacy frameworks that provide improved privacy estimates to overcome these limitations. This paper presents alpha differential privacy (ADP), a privacy architecture based on alpha divergence, a comprehensive set of metrics that assess the dissimilarity across probability distributions \cite{Amari2016, Csiszar2004}. This method provides a more flexible framework that may be customized for various application scenarios and sensitivity levels. ADP enhances the traditional differential privacy framework by leveraging the flexibility of alpha divergence, providing a range of privacy guarantees tailored to particular requirements.

Our research enhances the existing body of knowledge in multiple important aspects: we provide a robust theoretical framework for ADP, clarify its essential characteristics, and demonstrate its robustness and composability, comparable to other privacy frameworks. Empirical evaluations indicate that ADP offers significant advantages over several prominent privacy frameworks in contexts with a restricted number of iterations or strict privacy requirements, ADP attains a reduced initial privacy consumption with a decent privacy consumption growth in these contexts, underscoring its applicability in privacy-preserving data analysis, while also recognizing its limitations in other contexts. 

The organization of the subsequent sections of this paper is as follows: Section \MakeUppercase{\romannumeral 2} outlines differential privacy and its fundamental principles. Section \MakeUppercase{\romannumeral 3} examines the relevant literature on differential privacy and divergence metrics. Section \MakeUppercase{\romannumeral 4} provides a formal description of alpha differential privacy and examines its principal properties. Section \MakeUppercase{\romannumeral 5} examines the relationship between alpha differential privacy and approximate differential privacy. Section \MakeUppercase{\romannumeral 6} examines diverse mechanisms of alpha differential privacy, including experimental data and analysis that demonstrate its application and versatility.  Section \MakeUppercase{\romannumeral 7}  provides guidance on how to pick an appropriate $\alpha$ for a given task. Section \MakeUppercase{\romannumeral 8} presents the simulation settings, results, and an in-depth discussion. Section \MakeUppercase{\romannumeral 9} concludes the paper with a summary of findings and outlines prospective future research directions.

This work aims to advance the theoretical and practical understanding of alpha differential privacy, hence contributing to the development of more resilient and flexible privacy systems. This guarantees that privacy-preserving data analysis continues to be a feasible and powerful pursuit in a progressively data-driven world.

\section{Differential Privacy and Divergence}
Differential privacy (DP) was initially introduced by Dwork et al. \cite{Dwork2006_icalp} and has since established itself as the fundamental principle of privacy-preserving data analysis. It is a stringent mathematical framework that offers a formal notion of privacy for data analysis methods. The fundamental concept of differential privacy is to guarantee that the addition or removal of an individual from a dataset does not substantially influence the results of an algorithm. This inhibits an assailant from readily deducing any person's information.

\subsection{Pure Differential Privacy}
\begin{definition}[Pure differential privacy \cite{Dwork2014}] 
Let $\mathcal{D}$ denote a set of all possible datasets, and let $D_1, D_2 \in \mathcal{D}$ be two datasets that differ by exactly one element, denoted as $D_1 \sim D_2$. A randomized mechanism $\mathcal{M}$ that maps datasets to outputs in some range $\mathcal{R}$ is said to be pure differential privacy, or $\epsilon$-differential is defined as:

\begin{equation}
\Pr[\mathcal{M}(D_1) \in \mathcal{R}] \leq e^\epsilon \Pr[\mathcal{M}(D_2) \in \mathcal{R}].
\end{equation}
\end{definition}

The parameter $\epsilon$ regulates the degree of privacy. The selection of $\epsilon$ entails a compromise between privacy preservation and data utility. Lower values of $\epsilon$ enhance privacy guarantees but diminish the usefulness of the mechanism $\mathcal{M}$, while higher values of $\epsilon$ augment the utility of the mechanism $\mathcal{M}$ but compromise privacy safeguards. Standard values of $\epsilon$ vary from 0.01 to 1, contingent upon the particular privacy stipulations and objectives of the data analysis.

\subsection{Approximate Differential Privacy}
\begin{definition}[Approximate differential privacy \cite{dwork2006our}]
    Let $\mathcal{D}$ denote a set of all possible datasets, and let $D_1, D_2 \in \mathcal{D}$ be two datasets that differ by exactly one element, denoted as $D_1 \sim D_2$. A randomized mechanism $\mathcal{M}$ that maps datasets to outputs in some range $\mathcal{R}$ is said to be approximate differential privacy, or $(\epsilon, \delta)$-differential privacy if: 
\begin{equation}
\Pr\left( \Pr[\mathcal{M}(D_1) \in \mathcal{R}] > e^\epsilon \Pr[\mathcal{M}(D_2) \in \mathcal{R}] \right) \leq \delta.
\end{equation}
\end{definition}

Approximate differential privacy permits a minor failure possibility $\delta$ of breaching the privacy promise. This flexibility enables enhanced utility in data analytic jobs while preserving robust privacy safeguards. Approximate differential privacy is particularly advantageous in situations where the stringent condition of pure differential privacy ($\delta = 0$) is excessively limiting.

The inclusion of the $\delta$ parameter recognizes that in real-world applications, attaining complete privacy is frequently unfeasible or impracticable. By allowing a minimal risk of failure, $(\epsilon, \delta)$-differential privacy strikes a balance between privacy and usefulness, rendering it a flexible and extensively utilized privacy framework.

\subsection{Sensitivity and Noise Addition}

A key concept in designing differential private mechanisms is the sensitivity of a function $f$, defined as the maximum change in $f$'s output when its input dataset changes by one element. 
\begin{definition} [Sensitivity] For a function $f: \mathcal{D} \to \mathbb{R}^k$, the sensitivity $\Delta f$ is defined as:
\begin{equation}
\Delta f \triangleq \sup_{D_1, D_2: D_1 \sim D_2} \|f(D_1) - f(D_2)\|_p,
\end{equation}
where $\|\cdot\|_p$ denotes the $\ell_p$ norm. The choice of $p$ varies on the mechanism.
\end{definition}

To achieve differential privacy, noise proportional to the sensitivity of the function is added to the output. Common mechanisms include the Laplace mechanism and the Gaussian mechanism.

\subsubsection{Laplace Mechanism}  
For a function $f: D \to \mathbb{R}^k$, the Laplace mechanism $\mathcal{M}_L$ ensures $\epsilon$-differential privacy by adding noise drawn from the Laplace distribution to the output of the function \cite{Dwork2006}. Formally, the mechanism $\mathcal{M}_L$ is defined as:

\begin{equation}
\mathcal{M}_L(D) \triangleq f(D) + \text{Lap}(0,b),
\end{equation}
where $\text{Lap}(0,b)$ denotes the Laplace distribution with mean $0$ and scale parameter $b$. 

The Laplace mechanism ensures $(\epsilon, 0)$-differential privacy when:
\begin{equation}
    b = \frac{\Delta f_1}{\epsilon}
\end{equation}

Here, the sensitivity $\Delta f_1$ applies $\ell_1$ norm, which is the sum of the absolute differences between the corresponding elements of the vectors, this is also called $\ell_1$ sensitivity.

\begin{equation}
\Delta f_1 \triangleq \sup_{D_1, D_2: D_1 \sim D_2} \|f(D_1) - f(D_2)\|_1,
\end{equation}

\subsubsection{Gaussian Mechanism} For a function $f: D \to \mathcal{R}^k$, the Gaussian mechanism $\mathcal{M}_G$ ensures $(\epsilon, \delta)$-differential privacy by adding noise drawn from the Gaussian distribution to the output of the function \cite{dwork2006our}. Formally, the mechanism $\mathcal{M}_G$ is defined as:

\begin{equation}
\mathcal{M}_G(D) \triangleq f(D) + \mathcal{N}(0, \sigma^2),
\end{equation}
where $\mathcal{N}(0, \sigma^2)$ denotes the normal distribution with mean 0 and variance $\sigma^2$. 

The parameter $\sigma$ is chosen based on $\epsilon$, $\delta$, and $\Delta f$ to satisfy $(\epsilon, \delta)$-differential privacy:

\begin{equation}
\sigma^2 = \frac{2 \log(1.25/\delta) \Delta f_2^2}{\epsilon}.
\end{equation}

Here, the sensitivity $\Delta f_2$ of the function $f$ applies $\ell_2$ norm, this is also called $\ell_2$ sensitivity.

\begin{equation}
\Delta f_2 \triangleq \sup_{D_1, D_2: D_1 \sim D_2} \|f(D_1) - f(D_2)\|_2.
\end{equation}

The Gaussian mechanism is favoured for employing the $\ell_2$ norm as its measure of sensitivity. The $\ell_2$ norm consolidates the squared differences, which tends to mitigate bigger fluctuations, resulting in a diminished overall sensitivity score. Therefore, the noise introduced by the Gaussian mechanism is generally less intrusive than the noise necessitated by systems reliant on $\ell_1$ sensitivity, such as the Laplace mechanism. Furthermore, the Gaussian distribution's bell-shaped curve guarantees that the majority of the added noise is focused around the mean (zero), while the likelihood of extreme noise values decreases swiftly. This characteristic frequently yields outputs that approximate the actual function value more closely, enhancing the usefulness of the disseminated data.

\subsection{Divergence Measures}

Divergence measures are mathematical instruments employed to assess the disparity between probability distributions. Divergence metrics can be utilized within differential privacy to define and assess the privacy consumption resulting from a randomized method. Frequently utilized divergence metrics encompass Kullback–Leibler divergence, maximum divergence, and Rényi divergence.

\begin{definition}[Kullback–Leibler Divergence and Max Divergence]
Let $P$ and $Q$ be two probability measures over a measurable space $(\mathcal{X}, \mathcal{F})$ with property $P \ll Q$. By definition, \( P \ll Q \) (absolute continuity of \( P \) with respect to \( Q \)) means that for any measurable set \( A \in \mathcal{F} \), if \( Q(A) = 0 \), then \( P(A) = 0 \) as well. Throughout this paper, the notation \( \ll \) will consistently refer to absolute continuity.
The Kullback–Leibler (KL) divergence from $Q$ to $P$ is defined as:

\begin{align}
        D_{KL}(P \parallel Q) &\triangleq \int_{\mathcal{X}} \log \left( \frac{dP}{dQ} \right) \, dP \nonumber\\
        &= \mathbb{E}_P \left[\log \left( \frac{dP}{dQ} \right) \right].
\end{align}

The max divergence is defined as:

\begin{equation}
D_{\infty}(P \parallel Q) \triangleq \log (\operatorname*{ess\,sup}_{\mathcal{X}} \frac{dP}{dQ}),
\end{equation}
where $\frac{dP}{dQ}$ is the Radon-Nikodym derivative of $P$ with respect to $Q$, and $\mathbb{E}_P[\cdot]$ denotes the expectation with respect to the probability measure $P$.
\end{definition}
It is easy to see that the max divergence defined above is the worst-case analog of the KL divergence and it implies that the log-ratio of the probabilities is bounded by $\epsilon$, which directly relates to the max divergence:

\begin{equation}
D_{\infty}(\mathcal{M}(D) \parallel \mathcal{M}(D')) \leq \epsilon.
\end{equation}

Thus, the max divergence provides a useful and intuitive way to understand the worst-case privacy guarantees offered by a differential privacy mechanism \cite{Dwork2014}.

\begin{definition}[Rényi divergence] 
Let $P$ and $Q$ be two probability measures over a measurable space $(\mathcal{X}, \mathcal{F})$ with property $P \ll Q$.The Rényi divergence of order $\alpha$ between $P$ and $Q$ is defined as:

\begin{equation}
D_{\alpha}(P \parallel Q) = \frac{1}{\alpha - 1} \log \left[\int_{\mathcal{X}} \left(\frac{dP}{dQ}\right)^\alpha dQ\right],
\end{equation}
for $\alpha > 1$.
\end{definition}

\begin{lemma}
\begin{equation}
     \lim_{\alpha \to 1} D_{\alpha}(P \parallel Q) = D_{KL}(P \parallel Q). \label{eq:1}
\end{equation}
\begin{equation}
     \lim_{\alpha \to \infty} D_{\alpha}(P \parallel Q) = D_{\infty}(P \parallel Q). \label{eq:2}
\end{equation}
\end{lemma}

\begin{proof}
To prove
\begin{equation}
    \lim_{\alpha \to 1} D_{\alpha}(P \parallel Q) = D_{\mathrm{KL}}(P \parallel Q),
\end{equation}
define
\begin{equation}
    S(\alpha) = \int_{\mathcal{X}} P(x)^{\alpha} Q(x)^{1-\alpha} \, dx,
\end{equation}
and let \(f(\alpha) = \log S(\alpha)\). Then,
\begin{equation}
    D_{\alpha}(P \parallel Q) = \frac{f(\alpha)}{\alpha - 1}.
\end{equation}
Using differentiation under the integral sign:
\begin{equation}
    S'(\alpha) = \int_{\mathcal{X}} P(x)^{\alpha} Q(x)^{1-\alpha} (\log P(x) - \log Q(x)) \, dx.
\end{equation}
At \(\alpha = 1\), we have:
\begin{equation}
    S(1) = 1, \quad S'(1) = D_{\mathrm{KL}}(P \parallel Q).
\end{equation}
Thus,
\begin{equation}
    f'(1) = \frac{S'(1)}{S(1)} = D_{\mathrm{KL}}(P \parallel Q).
\end{equation}
Applying L'Hôpital's rule:
\begin{equation}
    \lim_{\alpha \to 1} D_{\alpha}(P \parallel Q) = \lim_{\alpha \to 1} \frac{f(\alpha)}{\alpha - 1} = f'(1) = D_{\mathrm{KL}}(P \parallel Q).
\end{equation}
Now, to prove
\begin{equation}
    \lim_{\alpha \to \infty} D_{\alpha}(P \parallel Q) = D_{\infty}(P \parallel Q),
\end{equation}
define
\begin{equation}
    I(\alpha) = \int_{\mathcal{X}} \left( \frac{P(x)}{Q(x)} \right)^{\alpha} Q(x) \, dx,
\end{equation}
such that
\begin{equation}
    D_{\alpha}(P \parallel Q) = \frac{1}{\alpha - 1} \log I(\alpha).
\end{equation}
Let \(M = \operatorname*{ess\,sup}_{x \in \mathcal{X}} \frac{P(x)}{Q(x)}\). For any \(\epsilon > 0\), define
\begin{equation}
    E_{\epsilon} = \left\{ x \in \mathcal{X} \, \bigg| \, \frac{P(x)}{Q(x)} > M - \epsilon \right\},
\end{equation}
where \(Q(E_{\epsilon}) > 0\) since \(M\) is the essential supremum.
We obtain the following bounds for \(I(\alpha)\):\\
\textit{Lower bound:}
\begin{equation}
    I(\alpha) \geq (M - \epsilon)^{\alpha} Q(E_{\epsilon}),
\end{equation}
\textit{Upper bound:}
\begin{equation}
    I(\alpha) \leq M^{\alpha}.
\end{equation}
Taking logarithms and dividing by \(\alpha\):
\begin{equation}
    \log(M - \epsilon) + \frac{1}{\alpha} \log Q(E_{\epsilon}) \leq \frac{\log I(\alpha)}{\alpha} \leq \log M.
\end{equation}
As \(\alpha \to \infty\), \(\frac{1}{\alpha} \log Q(E_{\epsilon}) \to 0\), so:
\begin{equation}
    \lim_{\alpha \to \infty} \frac{\log I(\alpha)}{\alpha} = \log M.
\end{equation}
Note that
\begin{equation}
    \lim_{\alpha \to \infty} \frac{\alpha}{\alpha - 1} = 1.
\end{equation}
Therefore,
\begin{align}
    \lim_{\alpha \to \infty} D_{\alpha}(P \parallel Q) &= \lim_{\alpha \to \infty} \frac{1}{\alpha - 1} \log I(\alpha) \nonumber\\
    &= \log M \nonumber\\
    &= D_{\infty}(P \parallel Q).
\end{align}
\end{proof}

\section{Related Work}
Given the properties outlined in Lemma 1, Rényi divergence is a natural choice for analyzing and developing new frameworks for differential privacy.

\subsection{Rényi differential privacy}
A direct application of Rényi divergence is Rényi Differential Privacy (RDP) \cite{Mironov2017}. RDP provides a more flexible and fine-grained privacy analysis compared to traditional differential privacy. A randomized mechanism $\mathcal{M}: D \to \mathbb{R}$ satisfies $(\alpha, \Bar{\epsilon})$-Rényi differential privacy if, for all adjacent datasets $D$ and $D'$,

\begin{equation}
D_{\alpha}(\mathcal{M}(D) \parallel \mathcal{M}(D')) \leq \Bar{\epsilon}.
\end{equation}

By adjusting $\alpha$, RDP allows precise control over the privacy-utility trade-off. Different $\alpha$ values provide varying sensitivity to outliers, enabling tailored privacy guarantees. RDP's strong composability properties simplify the analysis of cumulative privacy consumption. 

A common instantiation of the Rényi mechanism involves adding Gaussian noise. 
The parameter $\sigma_G^2$ is chosen to satisfy $(\alpha, \epsilon)$-RDP. Specifically, $\sigma_G^2$ is calibrated as:

\begin{equation}
\sigma_G^2 = \frac{ \alpha \Delta f_2^2}{2\Bar{\epsilon}}.
\end{equation}

Additionally, RDP can be converted to $(\epsilon, \delta)$-differential privacy, allowing for flexible privacy budget management. Specifically, given a mechanism that satisfies $(\alpha, \Bar{\epsilon})$-RDP. The parameter $\epsilon$ of $(\epsilon, \delta)$-differential privacy can be chosen as:

\begin{equation}
\epsilon = \Bar{\epsilon} + \frac{\log(1/\delta)}{\alpha - 1}.
\end{equation}

\subsection{Zero-Concentrated Differential Privacy}

Zero-Concentrated Differential Privacy (zCDP) is another refinement leveraging Rényi divergence \cite{bun2016concentrated}. A randomized mechanism $\mathcal{M}: D \to \mathbb{R}$ satisfies $\rho$-zCDP if for all adjacent datasets $D$ and $D'$ and for all $\alpha \in (1, \infty)$:

\begin{equation}
D_{\alpha}(\mathcal{M}(D) \parallel \mathcal{M}(D')) \leq \rho \alpha.
\end{equation}

zCDP simplifies privacy analysis compared to $(\Bar{\epsilon}, \delta)$-differential privacy. The Gaussian mechanism is a natural fit for zCDP, where for a function $f$ with $\ell_2$ sensitivity $\Delta f_2$, adding Gaussian noise with variance $\sigma_G^2$ satisfies $\rho$-zCDP with:

\begin{equation}
\sigma_G^2 = \frac{\Delta f_2^2}{2\rho}.
\end{equation}

zCDP also can be converted to $(\epsilon, \delta)$-differential privacy. Given a mechanism that satisfies $(\alpha, \Bar{\epsilon})$-zCDP. The parameter $\epsilon$ $(\epsilon, \delta)$-differential privacy can be chosen as:

\begin{equation}
\epsilon = \rho + 2 \sqrt{\rho\log(1/\delta)}.
\end{equation}

\section{Alpha differential Privacy}
Rényi divergence and Zero-Concentrated Differential Privacy offer robust and flexible frameworks for analyzing privacy consumption in differential privacy mechanisms. Their ability to balance privacy and utility, combined with strong composability properties, makes them essential tools in the design of privacy-preserving data analysis algorithms.

Inspired by Rényi divergence, this paper aims to explore the connection and practical significance of alpha divergence, a notable subset of f-divergence closely related to Rényi divergence, within the context of differential privacy \cite{villmann2010mathematical}.

\begin{definition}[$f$-divergence \cite{csiszar1967information}]
Given a convex function $f: (0, \infty) \rightarrow \mathbb{R}$ with $f(1) = 0$, if $P \ll Q$, the $f$-divergence between two probability measures $P$ and $Q$ over a measure space $(\mathcal{X}, \mathcal{F})$ is defined as:

\begin{equation}
    D_f(P \parallel Q) \triangleq \int_\mathcal{X} f\left(\frac{dP}{dQ}\right) dQ.
\end{equation}

In the context of probability density functions, let $\lambda$ be the Lebesgue measure, if $Q \ll \lambda$, the $f$-divergence $D_f(P \parallel Q)$ is just defined as:

\begin{equation}
    D_f(P \parallel Q) \triangleq \int_\mathcal{X} f\left[\frac{\left(\frac{dP}{d\lambda}\right)}{\left(\frac{dQ}{d\lambda}\right)}\right] {\frac{dQ}{d\lambda}}d\lambda = \int_\mathcal{X} f\left(\frac{p}{q}\right) q \, d\lambda.
\end{equation}

Where $p$ and $q$ are corresponding density functions of $P$ and $Q$ with respect to the Lebesgue measure (A more general version can be found in \cite{villmann2010mathematical}).

\end{definition}

\begin{definition}[Alpha divergence]
 Let $P$ and $Q$ be two probability measures over a measure space $(\mathcal{X}, \mathcal{F})$ and $\lambda$ be the Lebesgue measure with the property of $P \ll Q \ll \lambda$. Let $p$ and $q$ be the density functions of $P$ and $Q$ with respect to the Lebesgue measure. The alpha divergence is a special case of $f$-divergence, generated by the $f$-function defined on $\mathbb{R}\setminus \{0,1\}$ \cite{cichocki2008alpha}:

\begin{equation}
    f(u) = \frac{u^\alpha - \alpha u - (1-\alpha)}{\alpha(\alpha - 1)},
\end{equation}
where $u = \frac{dP}{dQ} = \frac{p}{q}$, reader can easily check the convexity of $f(u)$.\\
The alpha divergence can be expressed as:
\begin{align}
    \widetilde{D}_\alpha (P \parallel Q) &\triangleq \frac{1}{\alpha(\alpha-1)} \left[ \int_\mathcal{X} p^\alpha q^{1-\alpha} - \alpha p - (1-\alpha)q \, d\lambda\right] \notag\\
    &= \frac{1}{\alpha(\alpha-1)} \left[ \int_\mathcal{X} p^\alpha q^{1-\alpha} \, d\lambda - (1-\alpha) - \alpha\right] \notag \\
    &= \frac{1}{\alpha(\alpha-1)} \left[ \int_\mathcal{X} p^\alpha q^{1-\alpha} \, d\lambda - 1\right].
\end{align}
\end{definition}

\begin{lemma}
    An $f$-divergence is always non-negative.
\end{lemma}

\begin{proof}
Let $P$ and $Q$ be two probability measures over a measurable space $ (\mathcal{X}, \mathcal{F}) $. The $f$-divergence between $P$ and $Q$ is given by:

\begin{align}
        D_f(P \parallel Q) &= \int_\mathcal{X} f\left(\frac{dP}{dQ}\right) dQ \notag\\
        &= \mathbb{E}_{Q}\left[f\left(\frac{dP}{dQ}\right)\right].
\end{align}
Since $f$ is a convex function, by Jensen's inequality, for any random variable $X$,

\begin{equation}
    f(\mathbb{E}[X]) \leq \mathbb{E}[f(X)].
\end{equation}
Applying this to $X = \frac{dP}{dQ}$, we get:

\begin{equation}
    f\left(\mathbb{E}_{Q}\left[\frac{dP}{dQ}\right]\right) \leq \mathbb{E}_{Q}\left[f\left(\frac{dP}{dQ}\right)\right].
\end{equation}
Given that $\mathbb{E}_{Q}\left[\frac{dP}{dQ}\right] = 1$, it follows:

\begin{equation}
    f(1) = 0 \leq \mathbb{E}_{Q}\left[f\left(\frac{dP}{dQ}\right)\right].
\end{equation}
\end{proof}

Based on the non-negativity property of $f$-divergence (Lemma 2), we can define alpha differential privacy in a well-defined manner.

\begin{definition}[$(\alpha, \epsilon)$-ADP]
A randomized mechanism $\mathcal{M}: D \to \mathbb{R}$ satisfies $(\alpha, \epsilon)$-alpha differential privacy with $\alpha > 1$ if, for all adjacent datasets $D$ and $D'$,

\begin{equation}
\widetilde{D}_{\alpha}(\mathcal{M}(D) \parallel \mathcal{M}(D')) \leq \epsilon.
\end{equation}
\end{definition}
\begin{remark}
    Although $\alpha$ in alpha divergence can take any value except 0 and 1, ADP typically considers the case where $\alpha > 1$. This restriction is adopted because, in this range, ADP exhibits properties that are particularly advantageous for practical applications. These beneficial properties will be shown in detail below.
\end{remark}

\subsection{Preservation of alpha differential privacy under Post-Processing}

Differential privacy methods possess a crucial characteristic which is their capacity to tolerate post-processing, as stated in the works of \cite{Dwork2006_icalp}. This indicates that if a mechanism $\mathcal{M}$ satisfies the criteria for $\alpha$-differential privacy, then any additional operations performed on the output of $\mathcal{M}$ will not compromise the privacy safeguards that it provides. Due to the fact that it ensures that any further analysis or modification of data that has been anonymized by a differential privacy mechanism will not jeopardize the data's privacy, this trait is particularly significant in applications that are used in the real world.

Differential privacy is strong and highly relevant in a wide variety of data analytic workflows as a result of this resilience. One example of this would be in the field of machine learning, where a model could be trained on differential privacy data in order to prevent it from memorizing sensitive information. Following the completion of the training process, the model may undergo a number of evaluations and transformations, including parameter tuning and model compression, among others. The post-processing property ensures that the privacy protection that was initially provided to the training data will not be diminished as a result of these actions.

In the following, we present a substantial verification of this property, which demonstrates that ADP maintains its guarantees even after post-processing has been performed.

\begin{proposition}[Data Processing Inequality]
ADP is preserved under post-processing.
\end{proposition}

\begin{proof}
The proof follows the approach outlined in Erven's work \cite{erven2014renyi}. Let $ P $ and $ Q $ be two probability measures over a measurable space $ (\mathcal{X}, \mathcal{F}) $, with $ P \ll Q $. Let $ \mathcal{G} $ be the sub-$\sigma$-algebra of $ \mathcal{F} $ generated by a measurable map $ f $. We need to show that $ P_{\mathcal{G}} $ and $ Q_{\mathcal{G}} $, the restrictions of $ P $ and $ Q $ to $ \mathcal{G} $, satisfy
\begin{equation}
\widetilde{D}_{\alpha}(P_{\mathcal{G}} \parallel Q_{\mathcal{G}}) \leq \widetilde{D}_{\alpha}(P \parallel Q).
\end{equation}
To prove this, we need to show that:
\begin{equation}
\int_\mathcal{X} \left( \frac{dP_{\mathcal{G}}}{dQ_{\mathcal{G}}} \right)^{\alpha} dQ_{\mathcal{G}} \leq \int_\mathcal{X} \left( \frac{dP}{dQ} \right)^{\alpha} dQ.
\end{equation}
Recall that the conditional expectation $ \mathbb{E}_Q \left[ \frac{dP}{dQ} \bigg | \mathcal{G} \right] $ is the Radon-Nikodym derivative $ \frac{dP_{\mathcal{G}}}{dQ_{\mathcal{G}}} $.
\begin{equation}
\frac{dP_{\mathcal{G}}}{dQ_{\mathcal{G}}} = \mathbb{E}_Q \left[ \frac{dP}{dQ} \bigg| \mathcal{G} \right].
\end{equation}
By Jensen's inequality for the convex function $ x \mapsto x^{\alpha} $ (since by the definition of ADP, $ \alpha > 1 $), we have:
\begin{align}
    \int_\mathcal{X} \left( \frac{dP_{\mathcal{G}}}{dQ_{\mathcal{G}}} \right)^{\alpha} dQ_{\mathcal{G}} &= \int_\mathcal{X} \left( \frac{dP_{\mathcal{G}}}{dQ_{\mathcal{G}}} \right)^{\alpha} dQ \nonumber\\
       &= \int_\mathcal{X} \left( \mathbb{E} \left[ \frac{dP}{dQ} \bigg| \mathcal{G} \right] \right)^{\alpha} dQ \nonumber\\
      &\leq \int_\mathcal{X} \mathbb{E} \left[ \left( \frac{dP}{dQ} \right)^{\alpha} \bigg| \mathcal{G} \right] dQ \nonumber\\
      &= \int_\mathcal{X} \left( \frac{dP}{dQ} \right)^{\alpha} dQ.
\end{align}
The first line holds since \( dQ_{\mathcal{G}} \) is the restriction of \( dQ \) to the sub-\(\sigma\)-algebra \(\mathcal{G}\).
Hence, we have shown that:
\begin{equation}
\widetilde{D}_{\alpha}(P_{\mathcal{G}} \parallel Q_{\mathcal{G}}) \leq \widetilde{D}_{\alpha}(P \parallel Q).
\end{equation}
This proves that ADP is preserved under post-processing.
\end{proof}





\subsection{Adaptive composability of alpha differential privacy}
The ability of differential privacy methods to be composed is yet another important characteristic of these techniques. This composability guarantees that the cumulative privacy consumption that arises from numerous applications of differential privacy techniques may be controlled and regulated in a systematic manner, as stated in the works of\cite{Dwork2006, dwork2006our}. In order to demonstrate the efficacy and adaptability of ADP in comparison to general differential privacy methods, we will demonstrate that ADP possesses composability qualities that are comparable to those with general mechanisms. Furthermore, this promise extends to situations in which the next mechanism is picked in an adaptive manner based on the output of the mechanism that came before it. 

In machine learning pipelines, where models are frequently trained in an iterative manner, composability and adaptability are especially useful. It is possible that differential privacy safeguards will be implemented throughout each iteration. This will ensure that the model does not overfit the training data and so accidentally disclose sensitive information. ADP ensures that privacy guarantees are valid during the training process by retaining composability. This provides strong protection against data leakage and ensures that the training session is successful.

\begin{proposition}[Adaptive Sequential Composition] \label{adaptive}
Let mechanisms $\mathcal{M}_1: D \to \mathcal{A}$ and $\mathcal{M}_2: \mathcal{A} \times D \to \mathcal{B}$ be $(\alpha, \epsilon_1)$-ADP and $(\alpha, \epsilon_2)$-ADP mechanisms, respectively. Then the mechanism defined as $(X, Y)$, where $X \sim \mathcal{M}_1(D)$ and $Y \sim \mathcal{M}_2(X, D)$, satisfies $\left(\alpha, \epsilon_1 + \epsilon_2 + \alpha(\alpha - 1)\epsilon_1\epsilon_2 \right)$-ADP.
\end{proposition}

\begin{proof}
Let $\mathcal{M}_3: D \to \mathcal{A} \times \mathcal{B}$ be the mechanism obtained by sequentially applying $\mathcal{M}_1$ and $\mathcal{M}_2$. Denote the probability measures induced by $\mathcal{M}_1$ on $\mathcal{A}$, $\mathcal{M}_2$ given $X$ on $\mathcal{B}$, and the joint probability measure on $\mathcal{A} \times \mathcal{B}$ as $P_X$, $P_{Y|X}$, and $P_{X,Y}$, respectively. Similarly, let $P_{X'}$, $P_{Y'|X'}$, and $P_{X',Y'}$ represent the corresponding probability measures when the input dataset is $D'$. \\
Before proceeding with the calculations, note that \( P_{X, Y} \ll P_{X', Y'} \) holds because \( P_X \ll P_{X'} \) holds (from the ADP property of \( \mathcal{M}_1 \)) and \( P_{Y|X} \ll P_{Y'|X'} \) holds (from the ADP property of \( \mathcal{M}_2 \)). The product measure \( P_{X, Y} = P_X \times P_{Y|X} \) is therefore absolutely continuous with respect to \( P_{X', Y'} = P_{X'} \times P_{Y'|X'} \) given that all of the measures here is $\sigma$-finite. Then:
\begin{align}
    & \alpha(\alpha - 1) \widetilde{D}_{\alpha}(\mathcal{M}_3(D) \parallel \mathcal{M}_3(D')) + 1\nonumber\\
    = & \int_{\mathcal{A} \times \mathcal{B}} \left( \frac{dP_{X,Y}}{dP_{X',Y'}} \right)^{\alpha} dP_{X',Y'} \nonumber\\
    = & \int_{\mathcal{A} \times \mathcal{B}} \left( \frac{d(P_X  \times P_{Y|X})}{d(P_{X'}  \times P_{Y'|X'})} \right)^{\alpha} d(P_{X'}  \times P_{Y'|X'}) \nonumber\\
    = & \int_{\mathcal{A}} \left( \frac{dP_X}{dP_{X'}} \right)^{\alpha} dP_{X'} \int_{\mathcal{B}} \left( \frac{dP_{Y|X}}{dP_{Y'|X'}} \right)^{\alpha} dP_{Y'|X'} \nonumber\\
    \leq & \left( \alpha(\alpha - 1) \epsilon_1 + 1 \right) \left( \alpha(\alpha - 1) \epsilon_2 + 1 \right) \nonumber\\
    = & \alpha(\alpha - 1) \left( \epsilon_1 + \epsilon_2 + \alpha(\alpha - 1)\epsilon_1\epsilon_2 \right) + 1,
\end{align}
which proves the claim.
\end{proof}


 It should be noted that $\epsilon$ in alpha-differential privacy is not entirely consistent with the intuitive understanding of the privacy parameter. Higher values do not necessarily correspond to weaker privacy guarantees.  Since the mapping of parameters in alpha-differential privacy to those in traditional privacy frameworks is not linearly positively correlated. We will show this in section V.

\subsection{Group privacy of alpha differential privacy}
An extension of the individual privacy guarantees that are included in differential privacy frameworks is the concept of group privacy. Group privacy ensures that the privacy of any group of persons is also kept, in contrast to the standard differential privacy approach, which focuses on safeguarding the privacy of individual entries within a dataset \cite{mcsherry2009privacy}. In the context of ADP, the concept of group privacy addresses situations in which the adversary may possess auxiliary information about many persons contained within the dataset. This is of utmost significance in applications that deal with sensitive data, since it is necessary to safeguard the privacy of subgroups within the data against the possibility of inference attacks. 

Suppose that there is a database in the healthcare industry that contains confidential patient information. In the case of traditional differential privacy, it is possible to guarantee that the inclusion or exclusion of the data of a single patient does not significantly impact the outcomes of an analysis. On the other hand, if an adversary is aware that a group of patients are members of the same family or community, then they could be able to piece together sensitive information about the group by accessing data that is relevant to that group. Group privacy helps limit this danger by extending privacy guarantees to groups. This ensures that even if someone has access to additional information, they are unable to readily jeopardize the privacy of the individuals who are a part of the group.

The ADP that we have described can be extended to safeguard the privacy of groups of varying sizes. This will ensure that the appropriate level of privacy is provided for a variety of practical contexts, including healthcare, finance, social sciences, and other areas.

\begin{lemma}[Triangle inequality of alpha divergence]
Let $P$, $Q$, and $R$ be three probability measures defined in a measurable space $(\mathcal{X}, \mathcal{F})$, where $\lambda$ is the Lebesgue measure such that $\lambda \ll P \ll Q \ll R \ll \lambda$. Define $I_{\alpha}(A \parallel B) = \alpha(\alpha - 1) \widetilde{D}_{\alpha}(A \parallel B) + 1$, where $\widetilde{D}_{\alpha}(A \parallel B)$ denotes the alpha divergence between the probability measures $A$ and $B$ for $\alpha > 1$. Then, the following inequality holds:
\begin{equation*}
    I_{\alpha}(P \parallel Q) \leq \left(I_{2 \alpha}(P \parallel R)\right)^{\frac{1}{2}} \left(I_{2 \alpha - 1}(R \parallel Q)\right)^{\frac{1}{2}}.
\end{equation*}
\end{lemma}
\begin{proof}
The proof follows directly from an application of Hölder's inequality. We begin by expressing $I_{\alpha}(P \parallel Q)$ in terms of the Radon-Nikodym derivatives:

\begin{align}
    &I_{\alpha}(P \parallel Q) \nonumber\\
    &= \int_{\mathcal{X}} \left(\frac{dP}{dQ}\right)^{\alpha} dQ \nonumber\\
    &= \int_{\mathcal{X}} \left(\frac{dP}{d\lambda}\right)^{\alpha} \left(\frac{dR}{d\lambda}\right)^{\frac{1}{2} - \alpha} \left(\frac{dR}{d\lambda}\right)^{\alpha - \frac{1}{2}} \left(\frac{dQ}{d\lambda}\right)^{1 - \alpha} d\lambda.
\end{align}
Applying Hölder's inequality with exponents $p = 2$ and $q = 2$, we obtain:
\begin{align}
    I_{\alpha}(P \parallel Q) \nonumber&\leq \left( \int_{\mathcal{X}} \left(\frac{dP}{d\lambda}\right)^{2\alpha} \left(\frac{dR}{d\lambda}\right)^{1 - 2\alpha} d\lambda \right)^{\frac{1}{2}} \nonumber\\
    &\quad \times \left( \int_{\mathcal{X}} \left(\frac{dR}{d\lambda}\right)^{2\alpha - 1} \left(\frac{dQ}{d\lambda}\right)^{2 - 2\alpha} d\lambda \right)^{\frac{1}{2}}\nonumber\\
    &= \left( \int_{\mathcal{X}} \left(\frac{dP}{dR}\right)^{2\alpha} dR \right)^{\frac{1}{2}} \left( \int_{\mathcal{X}} \left(\frac{dR}{dQ}\right)^{2\alpha - 1} dQ \right)^{\frac{1}{2}} \nonumber\\
    &= \left( I_{2\alpha}(P \parallel R) \right)^{\frac{1}{2}} \left( I_{2\alpha - 1}(R \parallel Q) \right)^{\frac{1}{2}}.
\end{align}
This completes the proof.
\end{proof}

\begin{proposition}[Group privacy in ADP]
    Let $\mathcal{M}$ be a mechanism that satisfies $(\alpha, \epsilon)$-ADP with $\alpha > 2^k$. For any group of $2^k + 1$ sizes, let $D$ and $D'$ be two datasets that differ in at most $2^k$ entries. The mechanism $\mathcal{M}$ provides $\left(\frac{\alpha}{2^k}, \frac{\alpha(\alpha-1)}{\frac{\alpha}{2^k}\left(\frac{\alpha}{2^k}-1\right)}\epsilon\right)$-group privacy for any such pair of datasets $D$ and $D'$.
\end{proposition}
\begin{proof}
Let $D_1$, $D_2$, and $D_3$ be three datasets such that $D_1$ is adjacent to $D_2$ and $D_2$ is adjacent to $D_3$. Let $P$, $R$, and $Q$ be the probability measures induced by $\mathcal{M}(D_1)$, $\mathcal{M}(D_2)$, and $\mathcal{M}(D_3)$ over the measurable space $(\mathcal{X}, \mathcal{F})$, respectively. Assume \( \lambda \) is the Lebesgue measure, and suppose the condition \( \lambda \ll P \ll Q \ll R \ll \lambda \) is satisfied, as required by Lemma 3. Let \( I_{\alpha}(A \parallel B) = \alpha(\alpha - 1) \widetilde{D}_{\alpha}(A \parallel B) + 1 \), as defined in Lemma 3. Under these conditions, we can apply Lemma 3 to obtain:
\begin{equation}
    I_{\alpha}(P \parallel Q) \leq \left(I_{2 \alpha}(P \parallel R)\right)^{\frac{1}{2}} \left(I_{2 \alpha - 1}(R \parallel Q)\right)^{\frac{1}{2}}.
\end{equation}
Let's consider the relationship between $I_{2\alpha-1}(R \parallel Q)$ and $I_{2 \alpha}(R \parallel Q)$. We start with the following expression:
\begin{align}
    I_{2\alpha-1}(R \parallel Q) &= \mathbb{E}_Q\left[\left(\frac{dR}{dQ}\right)^{2 \alpha - 1}\right] \nonumber\\
    &= \mathbb{E}_Q\left[\left(\frac{dR}{dQ}\right)^{2 \alpha \cdot \frac{2 \alpha - 1}{2 \alpha}}\right].
\end{align}
By Jensen's inequality for the concave function $f(x) = x^{\frac{2 \alpha - 1}{2 \alpha}}$, we have:
\begin{equation}
    \mathbb{E}_Q\left[\left(\frac{dR}{dQ}\right)^{2 \alpha \cdot \frac{2 \alpha - 1}{2 \alpha}}\right] \leq I_{2\alpha}(R \parallel Q)^{\frac{2 \alpha - 1}{2 \alpha}}.
\end{equation}
Then,
\begin{align}
    I_{\alpha}(P \parallel Q) &\leq I_{2\alpha}(P \parallel R)^{\frac{1}{2}} I_{2\alpha}(R \parallel Q)^{\frac{2 \alpha - 1}{4 \alpha}} \nonumber\\
    &\leq I_{2\alpha}(P \parallel R),
\end{align}
which implies:
\begin{align}
    \widetilde{D}_{\frac{\alpha}{2}}(P \parallel Q) &\leq \frac{\alpha(\alpha - 1)}{\frac{\alpha}{2}\left(\frac{\alpha}{2} - 1\right)} \widetilde{D}_{\alpha}(P \parallel R) \notag\\
    &= \frac{\alpha(\alpha - 1)}{\frac{\alpha}{2}\left(\frac{\alpha}{2} - 1\right)} \epsilon.
\end{align}
Now, let \( D_1 \) and \( D_3 \) be two datasets differing in at most \( 2^k \) entries, where \( P \) and \( Q \) are the probability measures induced by \( \mathcal{M}(D_1) \) and \( \mathcal{M}(D_3) \), respectively. We maintain similar settings as before.
By induction, we have:
\begin{align}
    &\widetilde{D}_{\frac{\alpha}{2^k}}(P \parallel Q) \notag\\
    \leq& \frac{\frac{\alpha}{2^{k-1}}\left(\frac{\alpha}{2^{k-1}} - 1\right)} {\frac{\alpha}{2^{k}}\left(\frac{\alpha}{2^{k}} - 1\right)} \cdot \frac{\frac{\alpha}{2^{k-2}}\left(\frac{\alpha}{2^{k-2}} - 1\right)} {\frac{\alpha}{2^{k-1}}\left(\frac{\alpha}{2^{k-1}} - 1\right)} \cdots \frac{\alpha(\alpha - 1)}{\frac{\alpha}{2}\left(\frac{\alpha}{2} - 1\right)} \epsilon\notag\\
    =& \prod_{i=0}^{k-1} \frac{\frac{\alpha}{2^{i}}\left(\frac{\alpha}{2^{i}} - 1\right)} {\frac{\alpha}{2^{i+1}}\left(\frac{\alpha}{2^{i+1}} - 1\right)}\epsilon \notag\\
    =& \frac{\alpha(\alpha-1)}{\frac{\alpha}{2^k}\left(\frac{\alpha}{2^k}-1\right)}\epsilon,
\end{align}
which proves the claim.
\end{proof}

\begin{remark}
It is worth noting that maintaining a fixed \( \alpha \) while extending ADP to group privacy appears to be infeasible under the constraints of not assuming any specific distribution. Therefore, we must adjust \( \alpha \) by dividing it by \( 2^k \) to get the bound of group privacy.
\end{remark}

\begin{remark}
    Regarding the absolute continuity chain \( \lambda \ll P \ll Q \ll R \ll \lambda \), it is good to know that many commonly used differential privacy mechanisms, such as the Laplace and Gaussian mechanisms, satisfy this requirement. These mechanisms induce probability measures with well-defined Radon-Nikodym derivatives with respect to the Lebesgue measure \( \lambda \) (e.g., the Laplace and Gaussian densities), thereby ensuring absolute continuity. Additionally, these measures are mutually absolutely continuous because the mechanisms generate overlapping supports and assign nonzero probability density to the same regions of the space.
\end{remark}

\section{ADP and \texorpdfstring{$(\epsilon,\delta)$}{(epsilon，delta)}-DP}
The relationship between alpha differential privacy (ADP) and approximate differential privacy ($(\epsilon,\delta)$-DP) is an important aspect of privacy analysis. By adjusting the parameters $\alpha$ and $\epsilon$, ADP offers a flexible approach to privacy guarantees. This flexibility allows us to map the guarantees of ADP to the $( \epsilon, \delta)$ framework, thereby connecting these two important privacy models.

\begin{proposition}[Relationship between ADP and $(\epsilon,\delta)$-DP]
If $\mathcal{M}$ is an $(\alpha, \epsilon)$-ADP mechanism, then it also satisfies $(\Bar{\epsilon},\delta)$-DP with $\delta \in (0,1)$, where $\Bar{\epsilon} \geq \frac{\log\left(\frac{e^{\epsilon} \alpha(\alpha-1) + 1}{\delta}\right)}{\alpha-1}$. 
    
\end{proposition}
\begin{proof}
Let $P$ and $Q$ be the probability measures induced by $\mathcal{M}(D)$ and $\mathcal{M}(D')$, respectively, over a measurable space $(\mathcal{X}, \mathcal{F})$. We have:
\begin{align}
        \int_{\mathcal{X}} \left(\frac{dP}{dQ}\right)^\alpha \, dQ = &\mathbb{E}_Q \left[\left(\frac{dP}{dQ}\right)^\alpha\right] \nonumber\\
        = &\mathbb{E}_P \left[\left(\frac{dP}{dQ}\right)^{\alpha-1}\right] \nonumber\\
        \leq &\alpha(\alpha-1)\epsilon + 1.
\end{align}
By Markov’s inequality, we restrain:
\begin{align}
        \Pr\left[\frac{dP}{dQ} > e^{\Bar{\epsilon}}\right] &= \Pr\left[\left(\frac{dP}{dQ}\right)^{\alpha-1} \geq e^{\Bar{\epsilon}(\alpha-1)}\right] \nonumber\\
        &\leq \frac{\mathbb{E} \left[\left(\frac{dP}{dQ}\right)^{\alpha-1}\right]}{e^{\Bar{\epsilon}(\alpha-1)}} \nonumber\\
        &\leq \frac{\alpha(\alpha-1)\epsilon + 1}{e^{\Bar{\epsilon}(\alpha-1)}} \nonumber\\
        &\leq \delta,
\end{align}
which implies
\begin{equation}
\Bar{\epsilon} \geq \frac{\log\left(\frac{e^\epsilon \alpha(\alpha-1) + 1}{\delta}\right)}{\alpha-1}.
\end{equation}
This completes the proof.
\end{proof}

From the equation above, it is evident that in comparison to the parameter $\epsilon$, the approximate privacy guarantee corresponding to ADP is more constrained by the value of $\alpha$. Specifically, a larger $\alpha$ value tends to result in a more stringent approximate differential privacy guarantee.

In Section VII, we will conduct a detailed comparison of the privacy accumulation for various differential privacy.

\section{Various Mechanisms of ADP}
In this section, we explore three widely used mechanisms—Randomized response, Laplace, and Gaussian—and demonstrate how they can be adapted to the framework of alpha differential privacy (ADP). Each mechanism offers unique advantages and can be leveraged effectively depending on the specific requirements of a given privacy-preserving application.

\subsection{Randomized response mechanism}
The randomized response mechanism is frequently employed in privacy-preserving surveys and questionnaires. It incorporates randomization into responses to guarantee plausible deniability, complicating the identification of an individual's authentic response.

The mechanism $\mathcal{M}_R(f)$ for a predicate \( f: \mathcal{D} \to \{0,1\} \) is defined as follows:
\begin{equation}
    \mathcal{M}_R(f(D)) = \begin{cases} 
   f(D) & \text{with probability } p \\
   1 - f(D) & \text{with probability } 1 - p 
   \end{cases}
\end{equation}
Here, probability \( p \) controls the amount of noise introduced into the mechanism.
   \begin{proposition}[Randomized response mechanism and ADP]
If $\mathcal{M}_R$ is a randomized response mechanism, it satisfies $\left(\alpha,\frac{1}{\alpha(\alpha - 1)} \left( p^\alpha (1 - p)^{1 - \alpha} + (1 - p)^\alpha p^{1 - \alpha} - 1\right) \right)$-ADP. 
\end{proposition}

  \begin{proof}
   Without loss of generality, we assume that $f(D) = 1$ and the worst-case response generated by $D'$ is $f(D') = 0$.
   Using the definition of the randomized response mechanism, the probability distributions for $D$ and $D'$ are:
\begin{align}
   & Pr(\mathcal{M}_R(f(D)) = 1) = p,\ Pr(\mathcal{M}_R(f(D)) = 0) = 1 - p, \notag\\
    & Pr(\mathcal{M}_R(f(D')) = 1) = 1 - p,\ Pr(\mathcal{M}_R(f(D')) = 0) = p.
\end{align}
   Therefore,
   \begin{align}
            &\widetilde{D}_\alpha (\mathcal{M}_R(f(D)) \| \mathcal{M}_R(f(D'))) \nonumber\\
            = &\frac{1}{\alpha(\alpha-1)} \left( \sum_{\{0,1\}} p^\alpha q^{1-\alpha} - 1\right) \nonumber\\
            = &\frac{1}{\alpha(\alpha - 1)} \left( p^\alpha (1 - p)^{1 - \alpha} + (1 - p)^\alpha p^{1 - \alpha} - 1\right).
   \end{align}
  \end{proof}

  \begin{figure}[!htbp]
\centerline{\includegraphics[width=0.35\textwidth]{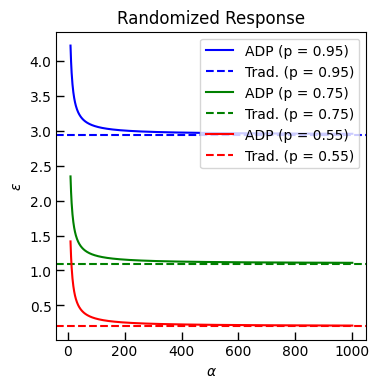}}
\caption{privacy consumption of the randomized response mechanism under different probabilities \( p \), with the horizontal axis representing the value of \(\alpha\) and the vertical axis indicating the privacy consumption (\(\epsilon\)). The solid line represents the privacy consumption evaluated by ADP for a failure probability of $\delta =$ 1$e$-5 across varying values of \(\alpha\), while the dashed line shows the privacy consumption as assessed by the traditional \(\epsilon\)-differential privacy framework.}
\label{fig:1}
\end{figure}

The randomized response mechanism stands out for its simplicity and applicability to categorical data, particularly in scenarios where the data consists of binary or discrete attributes. By flipping the result of a predicate with a certain probability, Randomized Response offers a straightforward yet effective way to ensure privacy while maintaining the utility of individual query results. As illustrated in Figure \ref{fig:1}, when ADP is used to evaluate privacy consumption of the randomized response mechanism, increasing the value of $\alpha$ causes the evaluation results to converge with those of the traditional privacy framework, which means this mechanism can be tightly integrated into the ADP framework, achieving a privacy-utility trade-off that through the choice of $\alpha$.

\subsection{Laplace mechanism}
  Recall that the definition of Laplace mechanism defined in \MakeUppercase{\romannumeral 2} is:
  \begin{equation}
\mathcal{M}_L(D) \triangleq f(D) + \text{Lap}(0,b),
\end{equation}

with $\ell_1$ sensitivity.

\begin{proposition}[Laplace mechanism and ADP]
If $\mathcal{M}_L$ is a Laplace mechanism, with sensitivity $\Delta f_1$ and scale $b$, it satisfies $\left(\alpha,  \frac{\exp\left(\frac{(\alpha-1)\mu}{b}\right)}{(\alpha-1)(2\alpha-1)} + \frac{\exp\left(-\frac{\alpha\mu}{b}\right)}{\alpha(2\alpha-1)} - \frac{1}{\alpha(\alpha-1)}\right)$-ADP. 
\end{proposition}

\begin{proof}
Without loss of generality, assume the distribution of $\mathcal{M}_L(D)$ is $\text{Lap}(0,b)$, the distribution $\mathcal{M}_L(D')$ generated by $D$'s adjacent dataset $D'$ is $\text{Lap}(\mu,b)$. Notice that the Laplace distribution is symmetrical. Thus, we can assume $\mu > 0$, we have:
\begin{align}
      &\widetilde{D}_\alpha (\mathcal{M}_L(D) \| \mathcal{M}_L(D')) \nonumber\\
        = &\frac{1}{\alpha(\alpha-1)} \frac{1}{2b} \bigg( \int_{-\infty}^0 \exp\left(\frac{x - \mu}{b}\right) \, dx \nonumber\\
        &\phantom{\frac{1}{\alpha(\alpha-1)} \frac{1}{2b}} + \int_{0}^{\mu} \exp\left(\frac{\alpha x - \mu}{b}\right) \, dx \nonumber\\
        &\phantom{\frac{1}{\alpha(\alpha-1)} \frac{1}{2b}} + \int_{\mu}^{+\infty} \exp\left(-\frac{\alpha x + \mu}{b}\right) \, dx \bigg) - 1 \nonumber\\
        = &\frac{1}{\alpha(\alpha-1)} \frac{1}{2b} \bigg(b\exp\left(\frac{(\alpha-1)\mu}{b}\right) \nonumber\\
        &\phantom{\frac{1}{\alpha(\alpha-1)} \frac{1}{2b}} + \frac{b}{2\alpha-1}\left(\exp\left(\frac{(\alpha-1)\mu}{b}\right) - \exp\left(-\frac{\alpha\mu}{b}\right)\right) \nonumber\\
        &\phantom{\frac{1}{\alpha(\alpha-1)} \frac{1}{2b}} + b\exp\left(-\frac{\alpha\mu}{b}\right) \bigg) - 1 \nonumber\\
        = &\frac{\exp\left(\frac{(\alpha-1)\mu}{b}\right)}{(\alpha-1)(2\alpha-1)} + \frac{\exp\left(-\frac{\alpha\mu}{b}\right)}{\alpha(2\alpha-1)} - \frac{1}{\alpha(\alpha-1)}.
\end{align}
For the multivariate Laplace mechanism, assume $\mu \in \mathbb{R}^d$, it is immediate that:
  \begin{align}
        & \widetilde{D}_\alpha (\mathcal{M}_L(D) \| \mathcal{M}_L(D')) \nonumber\\ 
          = &\frac{\exp\left(\frac{(\alpha-1)\|\mu\|_1}{b}\right)}{(\alpha-1)(2\alpha-1)} + \frac{\exp\left(-\frac{\alpha\|\mu\|_1}{b}\right)}{\alpha(2\alpha-1)} - \frac{1}{\alpha(\alpha-1)}.
  \end{align}

  We know that $\Delta f_1$ is an $\ell_1$ sensitivity. Therefore, we have:
\begin{equation}
    \Delta f_1 = \|\mu - 0\|_1 = \|\mu\|_1.
\end{equation}
This proves the claim.
\end{proof}

\begin{figure}[!htbp]
\centerline{\includegraphics[width=0.35\textwidth]{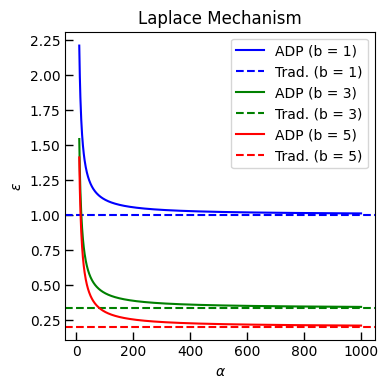}}
\caption{privacy consumption of the Laplace mechanism under different scales \( b \) and a fixed sensitivity $\Delta f_1 = 1$, with the horizontal axis representing the value of \(\alpha\) and the vertical axis indicating the privacy consumption (\(\epsilon\)). The solid line represents the privacy consumption evaluated by ADP for a failure probability of $\delta =$ 1$e$-5 across varying values of \(\alpha\), while the dashed line shows the privacy consumption as assessed by the traditional \(\epsilon\)-differential privacy framework.}
\label{fig:2}
\end{figure}

For the Laplace mechanism, the trend in privacy consumption (Figure \ref{fig:2}) closely mirrors that of the randomized response mechanism. As the value of alpha increases, the privacy consumption gradually converges to the results predicted by the traditional differential privacy framework.

\subsection{Gaussian Mechanism}
 Recall that the definition of Gaussian mechanism defined in \MakeUppercase{\romannumeral 2} is:
  \begin{equation}
\mathcal{M}_G(D) \triangleq f(D) + \mathcal{N}(0, \sigma_G^2),
\end{equation}

with $\ell_2$ sensitivity.

\begin{proposition}[Gaussian mechanism and ADP]\label{Gaussian}
If $\mathcal{M}_G$ is a Gaussian mechanism, with sensitivity $\Delta f_2$ and variance $\sigma_G^2$, it satisfies $\left(\alpha,\frac{1}{\alpha(\alpha-1)} \left( \exp{\left(\frac{(\alpha^2 - \alpha)\Delta f_2^2}{2\sigma_G^2}\right)}- 1\right)\right)$-ADP. 
\end{proposition}

\begin{proof}
Similar to the proof of the Laplace mechanism. Without loss of generality, assume the distribution of $\mathcal{M}_G(D)$ is $\mathcal{N}(0,\sigma_G^2)$, the distribution of $\mathcal{M}_G(D')$ generated by $D$'s adjacent dataset $D'$ is $\mathcal{N}(\mu,\sigma_G^2)$. Hence, we have:
\begin{align}
        &\widetilde{D}_\alpha (\mathcal{M}_G(D) \| \mathcal{M}_G(D')) \nonumber\\
        = &\frac{1}{\alpha(\alpha-1)} \int_{-\infty}^{+\infty} \frac{\exp{\left(\frac{-\alpha x^2-(1-\alpha) (x-\mu)^2}{2\sigma_G^2}\right)}}{\sigma_G\sqrt{2\pi}}- 1 \nonumber\\
         = &\frac{1}{\alpha(\alpha-1)} \left( \exp{\left(\frac{(\alpha^2 - \alpha)\mu^2}{2\sigma_G^2}\right)}- 1\right).
\end{align}
For the multivariate Gaussian mechanism, assume $\mu \in \mathbb{R}^d$, then, the distribution of $\mathcal{M}_G(D)$ is $\mathcal{N}(0,\sigma_G^2I_d)$, the distribution of $\mathcal{M}_G(D')$ is $\mathcal{N}(\mu,\sigma_G^2I_d)$, it is immediate that:
\begin{align}
        &\widetilde{D}_\alpha (\mathcal{M}_G(D) \| \mathcal{M}_G(D')) \nonumber\\
          = &\frac{1}{\alpha(\alpha-1)} \left( \exp{\left(\frac{(\alpha^2 - \alpha)\|\mu\|_2^2}{2\sigma_G^2}\right)}- 1\right).
\end{align}
We know that $\Delta f_2$ is an $\ell_2$ sensitivity. Therefore, we have:
\begin{equation}
    \Delta f_2^2 = \|\mu - 0\|_2^2 = \|\mu\|_2^2.
\end{equation}
This proves the claim.
\end{proof}

\begin{corollary}
    A Gaussian mechanism with variance $\frac{\alpha(\alpha-1)\Delta f_2^2}{2\log{\left(\alpha(\alpha-1)\epsilon+1\right)}}$ satisfies $(\alpha,\epsilon)$-ADP
\end{corollary}
\begin{proof}
    The proof is immediate from Proposition \ref{Gaussian}.
\end{proof}

\begin{figure}[!htbp]
\centerline{\includegraphics[width=0.35\textwidth]{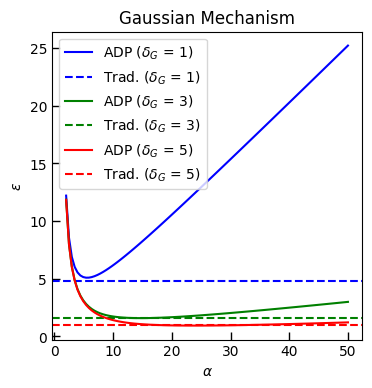}}
\caption{privacy consumption of the Gaussian mechanism under different standard deviation $\sigma_G$ and a fixed sensitivity $\Delta f_2 = 1$, with the horizontal axis representing the value of \(\alpha\) and the vertical axis indicating the privacy consumption (\(\epsilon\)). The solid line represents the privacy consumption evaluated by ADP for a failure probability of $\delta =$ 1$e$-5 across varying values of \(\alpha\), while the dashed line shows the privacy consumption as assessed by the traditional $(\epsilon,\delta)-$differential privacy framework.}
\label{fig:3}
\end{figure}

Under the ADP framework, the privacy consumption trend of the Gaussian mechanism displays a unique pattern in contrast to the preceding two mechanisms. Figure \ref{fig:3} demonstrates that the privacy consumption initially exhibits a convergence pattern as the amount of $\alpha$ increases, closely aligning with the privacy bounds anticipated by conventional differential privacy. This convergence is transient; once reaching a specific threshold, the privacy consumption diverges from the standard of traditional differential privacy. This trend is especially pronounced when the variance parameter $\sigma_G$ is minimal (e.g., $\sigma_G = 1$), as the privacy consumption under ADP markedly surpasses that of the conventional framework with increasing $\alpha$. This non-monotonic behaviour underscores the intricate relationship between $\alpha$ and privacy assurances in ADP, indicating that the selection of an optimal $\alpha$ necessitates meticulous evaluation, as an inappropriate choice may lead to greater cumulative privacy consumption than conventional privacy methods.

The strength of the ADP framework does not lie in evaluating privacy consumption for a single query. Instead, its real advantage lies in providing an effective upper bound estimation for the cumulative privacy consumption across multiple iterations, which will be analyzed in detail in Section  \MakeUppercase{\romannumeral 8}.

\section{Guidance on Choosing \texorpdfstring{$\alpha$}{alpha}}
In actual applications that require multiple iterations.  A carefully chosen $\alpha$ ensures the privacy budget is utilized efficiently, minimizing the cumulative privacy loss. This section takes the Gaussian mechanism as an example to show how to select the optimal $\alpha$.
\begin{figure}[!htbp]
\centerline{\includegraphics[width=0.35\textwidth]{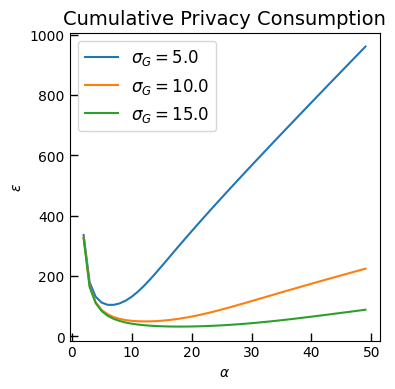}}
\caption{Cumulative privacy consumption (\(\epsilon\)) of the Gaussian mechanism over 1000 iterations under varying required standard deviations, with a required failure rate of \(\delta =\)1$e$-5 and a required sensitivity of \(\Delta f_2 = 1\). The horizontal axis represents the value of \(\alpha\), while the vertical axis indicates the cumulative privacy consumption (\(\epsilon\)).}
\label{fig:privacy_vs_alpha}
\end{figure}

Figure \ref{fig:privacy_vs_alpha} illustrates the relationship between the cumulative privacy consumption of a Gaussian mechanism over 1000 iterations and \(\alpha\). The privacy consumption function exhibits a convex-like behaviour: for small values of \(\alpha\), privacy consumption decreases rapidly as \(\alpha\) increases, reaching a minimum within an optimal range. Beyond this range, further increases in \(\alpha\) result in a gradual rise in privacy consumption. Notably, this trend closely resembles that of a single query using the Gaussian mechanism, as depicted in Figure \ref{fig:3}.

Similar to \(\alpha\) in Rényi Differential Privacy, \(\alpha\) in ADP is dynamically determined based on user-defined constraints. Algorithm \ref{alg:find_optimal_alpha_adp} demonstrates a scenario on how to determine the optimal \(\alpha\) to minimize cumulative privacy consumption based on user-defined constraints, including the failure probability, total number of iterations, standard deviation, and sensitivity. In this algorithm, \(\alpha\) is dynamically selected to minimize cumulative privacy consumption under specific requirements. 

\begin{figure}[!htbp]
\centerline{\includegraphics[width=0.35\textwidth]{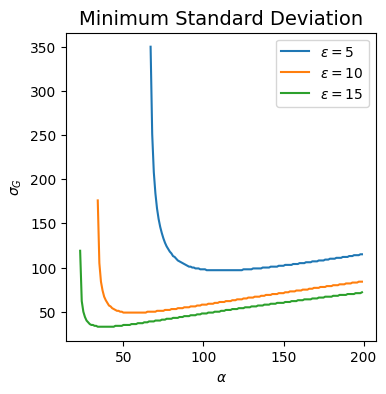}}
\caption{Optimal standard deviation (\(\sigma_G\)) of the Gaussian mechanism over 1000 iterations under varying required upper bounds of cumulative privacy consumption, with a required failure rate of \(\delta =\) 1$e$-5 and a required sensitivity of \(\Delta f_2 = 1\). The horizontal axis represents the value of \(\alpha\), while the vertical axis indicates the standard deviation (\(\sigma_G\)).}
\label{fig:std_vs_alpha}
\end{figure}

\begin{algorithm}
\caption{Find Alpha that Minimizes Privacy Consumption for Gaussian Mechanism}
\label{alg:find_optimal_alpha_adp}
\begin{algorithmic}[1]
 \renewcommand{\algorithmicrequire}{\textbf{Input:}}
 \renewcommand{\algorithmicensure}{\textbf{Output:}}
 \newcommand{\algcomment}[1]{\hfill \text{// #1}}
 \REQUIRE Number of iterations $\ell$, Standard deviation $\sigma_G$, Failure probability $\delta$, Sensitivity $\Delta f_2$\\
 \ENSURE  Minimum privacy consumption $\epsilon_{min}$ and optimal alpha $\alpha^*$
\STATE Initialize $\epsilon_{min} \gets \infty$, $\alpha^* \gets 2$\\
\STATE \text{//} E.g. we can take $\alpha$ from 2 to 100\\
\FOR{$\alpha$ in a suitable range}
    \STATE Compute the ADP privacy consumption for a single query.\\
    \STATE $\epsilon \gets \frac{1}{\alpha (\alpha - 1)} \left( \exp \left( \frac{(\alpha^2 - \alpha)\Delta f_2^2}{2 \sigma_G^2} \right) - 1 \right)$\\
    \STATE Compute the cumulative ADP privacy consumption for $\ell$ iterations.
    \STATE $\epsilon_{new}\gets 0$
    \FOR{$i = 1$ to $\ell$}
        \STATE Update $\epsilon_{new} \gets \epsilon_{new}  +  \epsilon + \alpha (\alpha - 1) \cdot  \epsilon \cdot \epsilon_{new} $
    \ENDFOR
    \STATE Convert ADP privacy consumption to traditional privacy consumption.\\
    \STATE $\epsilon_{temp} \gets \frac{\log \left( \epsilon_{new} \cdot \alpha (\alpha - 1) + 1 \right)}{\delta (\alpha - 1)}$\\
    \IF{$\epsilon_{temp} < \epsilon_{min}$}
        \STATE Update $\epsilon_{min} \gets \epsilon_{temp}$, $\alpha^* \gets \alpha$
     \ENDIF
\ENDFOR
\STATE \textbf{Return} $\epsilon_{min}$, $\alpha^*$
\end{algorithmic}
\end{algorithm}
For scenarios where users have different requirements—such as minimizing the standard deviation of the Gaussian mechanism to enhance data utility—a similar algorithm can be employed (Algorithm \ref{alg:find_optimal_std}). By leveraging the specified failure probability, total iterations, sensitivity, and an upper bound on the overall privacy consumption, the optimal \(\alpha\) and its corresponding standard deviation can be effectively determined. As shown in Figure \ref{fig:std_vs_alpha}.

\begin{algorithm}
\caption{Find Alpha that Minimizes Standard Deviation for Gaussian Mechanism}
\label{alg:find_optimal_std}
\begin{algorithmic}[1]
 \renewcommand{\algorithmicrequire}{\textbf{Input:}}
 \renewcommand{\algorithmicensure}{\textbf{Output:}}
 \newcommand{\algcomment}[1]{\hfill \text{// #1}}
 \REQUIRE Number of iterations $\ell$, Privacy consumption bound $\epsilon_{bound}$, Failure probability $\delta$, Sensitivity $\Delta f_2$\\
 \ENSURE  Minimum standard deviation $\sigma_{min}$ and optimal alpha $\alpha^*$
\STATE Initialize $\alpha^* \gets 2$, $\sigma_{min}\gets \infty$\\
\STATE \text{//} E.g. we can take $\alpha$ from 2 to 100\\
\FOR{$\alpha$ in a suitable range}
\STATE \text{//} E.g. we can take $\sigma_G$ from 1 to 500\\
    \FOR{$\sigma_G$ in a suitable range}
    \IF{$\sigma_{G} \geq \sigma_{min}$}
        \STATE \textbf{break}
    \ENDIF
    \STATE Compute the ADP privacy consumption for a single query.\\
    \STATE $\epsilon \gets \frac{1}{\alpha (\alpha - 1)} \left( \exp \left( \frac{(\alpha^2 - \alpha)\Delta f_2^2}{2 \sigma_G^2} \right) - 1 \right)$\\
    \STATE Compute the cumulative ADP privacy consumption for $\ell$ iterations.
    \STATE $\epsilon_{new}\gets 0$
    \FOR{$i = 1$ to $\ell$}
        \STATE Update $\epsilon_{new} \gets \epsilon_{new}  +  \epsilon + \alpha (\alpha - 1) \cdot  \epsilon \cdot \epsilon_{new} $
    \ENDFOR
    \STATE Convert ADP privacy consumption to traditional privacy consumption.\\
    \STATE $\epsilon_{temp} \gets \frac{\log \left( \epsilon_{new} \cdot \alpha (\alpha - 1) + 1 \right)}{\delta (\alpha - 1)}$\\
    \IF{$\epsilon_{temp} < \epsilon_{bound}$ }
        \STATE Update $\sigma_{min} \gets \sigma_{G}$, $\alpha^* \gets \alpha$
        \STATE \textbf{break}
     \ENDIF
\ENDFOR
\ENDFOR
\STATE \textbf{Return} $\sigma_{min}$, $\alpha^*$
\end{algorithmic}
\end{algorithm}

In conclusion, our empirical results show that, over multiple iterations, the relationship between \(\alpha\) and other privacy parameters is non-monotonic, instead exhibiting a behaviour that resembles convexity. This means that excessively high or low values of \(\alpha\) can adversely impact the dependent privacy parameters, thereby affecting overall performance. This highlights the necessity of dynamically selecting \(\alpha\) based on specific constraints to optimize privacy requirements or data utility. Our simulations indicate that evaluating a small range of \(\alpha\) values—typically between 2 and 300—is generally sufficient to identify the optimal choice.

\section{Simulation and Discussion}
In this section, we conduct simulations to explore the privacy consumption of different differential privacy frameworks, including alpha differential privacy (ADP), Rényi Differential Privacy (RDP), Zero-Concentrated Differential Privacy (zCDP), and the Advanced Composition (Adv.) theorem \cite{Dwork2010Boosting} under different iteration scenarios. This section is further divided into three parts: the simulation settings, the obtained results, and a detailed discussion of the observed trends.

\subsection{Simulation Settings}
Our simulations concentrate on the privacy consumption of the Gaussian mechanism, which is extensively employed in numerous differential privacy applications owing to its advantageous characteristics, especially in maintaining the utility of processed data. We choose the results generated by Gaussian as the basis for comparison to ensure that the results reflect real-world settings where different privacy approaches are often adopted. The simulations adjust the number of iterations and the failure probability $\delta$, to compare the performance of each mechanism under different conditions, ranging from a minimum to a wide range of iterations and varying $\delta$ values. We also establish the variance parameter $\sigma_G$ at various levels ($\sigma_G = 10, 50, 100$) to assess the sensitivity of each privacy framework to this variable. For alpha differential privacy (ADP) and Rényi differential privacy (RDP), we demonstrate the results under their optimal parameter choice for $\alpha$, providing an evaluation of their performance when optimally configured. It should be noted that for the curves of Adv., \( \delta \) represents the overall \( \delta \) after applying the advanced composition theorem. 


For Figure \ref{fig:4}, the $\epsilon$ values for a single query under the ADP framework are 5.00$e$-5, 5.05$e$-5, and 5.24$e$-5, respectively. Similarly, for Figure \ref{fig:5}, the $\epsilon$ values for a single query under ADP are 5.29$e$-5, 5.19$e$-5, and 5.12$e$-5, respectively. For Figure \ref{fig:6}, the $\epsilon$ values for a single query under ADP are 5.02$e$-3, 2.01$e$-4, and 5.04$e$-5, respectively. It is important to emphasize that these $\epsilon$ values represent the privacy parameter $\epsilon$ defined within the ADP frameworks, rather than the privacy consumption in the traditional $(\epsilon, \delta)$-differential privacy framework mentioned earlier. These data are provided here for the readers' reference.

\subsection{Simulation Results}
\begin{figure*}[!htbp]
\centerline{\includegraphics[width=\textwidth]{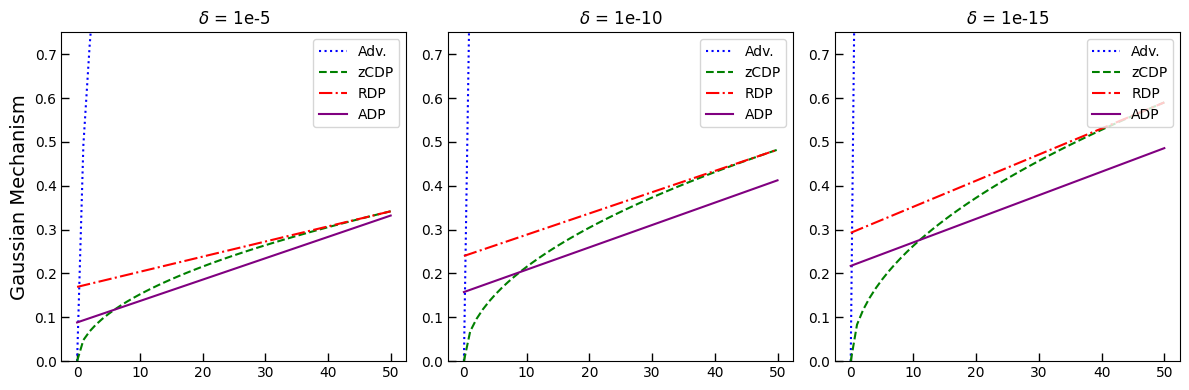}}
\caption{Comparison of cumulative privacy consumption between alpha differential privacy and other mainstream differential privacy frameworks for the Gaussian mechanism under small iterations, with a fixed standard deviation $\sigma_G = 100$ and a fixed sensitivity $\Delta f_2 = 1$. The horizontal axis represents the number of iterations, and the vertical axis represents the corresponding privacy consumption $(\epsilon)$. The blue line represents the advanced composition of differential privacy, the red line represents Rényi differential privacy, the green line represents zero-concentrated differential privacy, and the purple line represents alpha differential privacy. The results are shown for three different values of the failure probability: $\delta = $1$e$-5, $\delta = $1$e$-10, and $\delta = $1$e$-15. For ADP, the selected $\alpha$ values to minimize cumulative privacy consumption are $136$, $152$, and $164$, respectively, while for RDP, the corresponding 
$\alpha$ values are $69$, $97$, and $119$.}
    \label{fig:4}
\end{figure*}

\begin{figure*}[!htbp]
\centerline{\includegraphics[width=\textwidth]{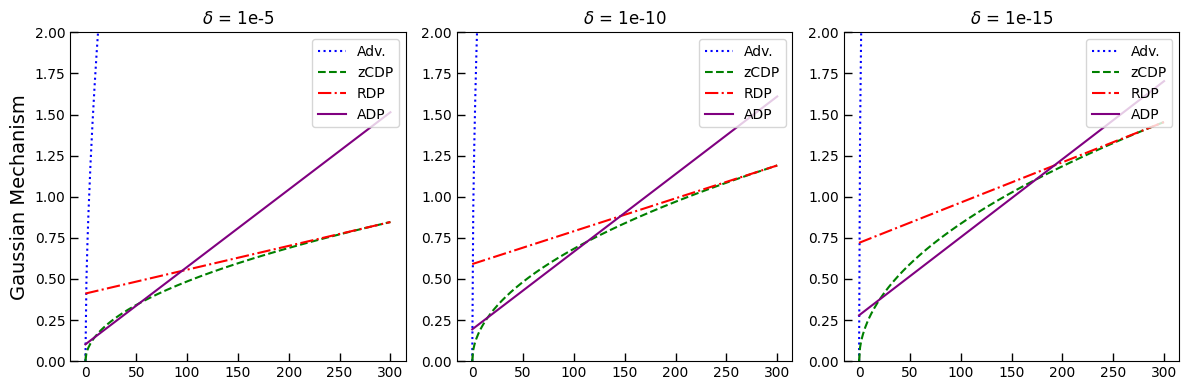}}
\caption{Comparison of cumulative privacy consumption between alpha differential privacy and other mainstream differential privacy frameworks for the Gaussian mechanism under relatively large iterations, with a fixed standard deviation $\sigma_G = 100$ and a fixed sensitivity $\Delta f_2 = 1$. The horizontal axis represents the number of iterations, and the vertical axis represents the corresponding privacy consumption $(\epsilon)$. The blue line represents the advanced composition of differential privacy, the red line represents Rényi differential privacy, the green line represents zero-concentrated differential privacy, and the purple line represents alpha differential privacy. The results are shown for three different values of the failure probability $\delta =$1$e$-5, $\delta =$1$e$-10, and $\delta =$1$e$-15. For ADP, the selected $\alpha$ values to minimize cumulative privacy consumption are $13$, $64$, and $127$, respectively, while for RDP, the corresponding 
$\alpha$ values are $6$, $25$, and $49$.}
    \label{fig:5}
\end{figure*}

\begin{figure*}[!htbp]
    \centering
    \includegraphics[width=\textwidth]{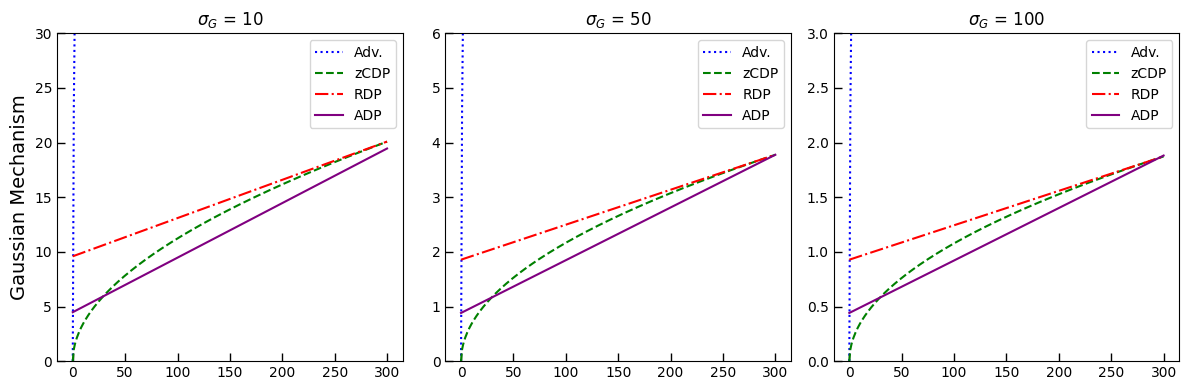}
    \caption{Comparison of cumulative privacy consumption between alpha differential privacy and other mainstream differential privacy frameworks for the Gaussian mechanism under relatively large iterations, with a fixed failure probability $\delta = $1$e$-25 and a fixed sensitivity $\Delta f_2 = 1$. The horizontal axis represents the number of iterations, and the vertical axis represents the corresponding privacy consumption $(\epsilon)$. The blue line represents the advanced composition of differential privacy, the red line represents Rényi differential privacy, the green line represents zero-concentrated differential privacy, and the purple line represents alpha differential privacy. The results are shown for three different values of the standard deviation $\sigma_G = 10$, $\sigma_G = 50$, and $\sigma_G = 100$. For ADP, the selected $\alpha$ values to minimize cumulative privacy consumption are $14$, $67$, and $133$, respectively, while for RDP, the corresponding 
$\alpha$ values are $7$, $32$, and $63$.}
    \label{fig:6}
\end{figure*}

Our simulation results are shown in Figures \ref{fig:4}, \ref{fig:5}, and \ref{fig:6}, which compare the privacy consumption of different differential privacy mechanisms, including ADP, RDP, zCDP, and Advanced Composition, in detail. These figures aim to illustrate the effectiveness of each privacy framework under varying conditions, such as different numbers of iterations and different values of the failure probability $\delta$. In these figures, the horizontal axis represents the number of iterations, while the vertical axis shows the corresponding privacy consumption. This visualization allows for a detailed comparison of how each mechanism performs in terms of cumulative privacy consumption over multiple iterations. 

Figure \ref{fig:4} shows the privacy consumption trends of different mechanisms under three different $\delta$ values (1e-5, 1e-10, and 1e-15) with a small number of iterations. The main observation is that although the privacy consumption of both ADP and RDP estimates shows a linear growth, the lower intercept of ADP shows that it provides a stronger initial privacy estimate. 

Figure \ref{fig:5} shows the privacy consumption when the number of iterations is relatively large, where $\delta$ is set to 1$e$-5. Here, we observe the performance difference between ADP and other frameworks as the number of iterations increases. While ADP starts with a relatively lower privacy consumption, it has a steeper slope compared to RDP, increasing the cumulative privacy consumption as the number of iterations increases.


Figure \ref{fig:6} provides the privacy consumption under strict failure probability requirements ($\delta = $1$e$-25). In this case, ADP maintains a clear advantage over RDP regarding the increasing slope of the privacy consumption curve, indicating that its cumulative privacy consumption grows slower than RDP as $\delta$ becomes smaller.

Figures \ref{fig:4}, \ref{fig:5}, and \ref{fig:6} illustrate that although zCDP initially exhibits a higher growth rate of privacy consumption, its logarithmic growth becomes favourable over a wide range of iterative scenarios.

Each figure also highlights the consistent behaviour of the advanced combination mechanism, which shows higher privacy consumption in all scenarios compared to ADP, RDP, and zCDP. 

\subsection{Discussion}
One of the key observations from the results is the behaviour of ADP versus RDP under small iterations. Figure \ref{fig:4} illustrates that both ADP and RDP demonstrate linear growth in privacy consumption. Nonetheless, ADP has a continually lower intercept, signifying a diminished initial privacy consumption. This attribute indicates that ADP is very efficient in situations necessitating a limited number of repeats, hence providing enhanced initial privacy assurances relative to RDP. However, this initial benefit gradually diminishes as the number of iterations increases, with ADP having a higher growth rate of privacy consumption than RDP, ultimately leading to a larger cumulative privacy consumption over a large number of iterations. This trade-off must be meticulously evaluated when choosing a differential privacy method, especially for applications that entail repetitive queries.

Another important finding in the results is related to zCDP. As the number of iterations increases, its logarithmic growth rate under the combination becomes increasingly favourable. The continued decline in the growth rate of zCDP with increasing iterations allows it to remain stable across various privacy settings. Despite the high initial growth rate of privacy consumption, zCDP is well suited for situations where data needs to be accessed frequently or for long-term continuous analysis, where managing the cumulative privacy consumption is essential.

The findings also underscore a notable aspect of ADP under rigorous $\delta$ criteria, as illustrated in Figure \ref{fig:6}. With a tight failure probability promise ($\delta=$ 1$e$-25), the slope of the privacy consumption curve for ADP increases at a slower rate than that of RDP, indicating that ADP is especially appropriate for situations necessitating exceptionally rigorous privacy assurances. This attribute renders ADP beneficial for applications dealing with extremely sensitive data, where minimizing privacy consumption during repeated accesses is essential.

In contrast, the advanced composition framework, indicated by the blue line in all figures, consistently demonstrates the largest privacy consumption in every scenario. This persistently elevated expense constrains its applicability in contexts where reducing privacy consumption is a primary goal. The advanced composition approach may remain relevant in situations when simpler privacy accounting is favoured and computing speed is emphasized over the reduction of cumulative privacy consumption.

Observations above demonstrate that ADP exhibits compelling advantages in practical scenarios characterized by small to moderate iterations and stringent failure probability requirements (\(\delta\)). These conditions are particularly prevalent in highly sensitive domains such as healthcare and finance, where robust privacy guarantees are imperative. In healthcare applications, particularly electronic health record (EHR) analysis, privacy regulations such as 
Health Insurance Portability and Accountability Act (HIPAA) mandate extraordinarily stringent privacy safeguards. Typical scenarios involve constrained query patterns (approximately less than 100 iterations) with extremely low failure probabilities (e.g., $\delta=$ 1$e$-15 or smaller) to protect sensitive patient information. In such cases, ADP enables dynamic selection of \(\alpha\) to minimize cumulative privacy consumption while meeting the given constraints. Our empirical analysis, as illustrated in Figure \ref{fig:4}, demonstrates that with $\delta=$ 1$e$-15 and 50 iterations, ADP achieves approximately $20\%$ reduction in cumulative privacy consumption compared to existing frameworks like RDP and zCDP by determining an optimal \(\alpha\).

The advantages of ADP extend similarly to financial applications, where protecting sensitive financial data (e.g., account transactions, credit histories, investment portfolios) is crucial. In scenarios such as credit risk assessment and fraud detection systems, which typically require 50 to 200 iterations, ADP's adaptive framework demonstrates superior performance. As evidenced in Figure \ref{fig:6}, under the extreme constraint of failure probability ($\delta=$ 1$e$-25), ADP outperforms other privacy frameworks in minimizing cumulative privacy consumption. These scenarios highlight the practical significance of ADP in real-world applications.

\section{Conclusion and Future Work}
This section concludes the findings of our research and outlines potential directions for future work.

\subsection{Conclusion}
The results of this study demonstrate that alpha differential privacy (ADP) is particularly appropriate for applications with small to moderate iterations, especially in settings where the failure probability needs to be strictly limited. Alpha divergence provides ADP with the necessary flexibility to fine-tune privacy consumption while achieving a customized balance between privacy and utility. In the small iteration setting, ADP has a unique advantage in that it can evaluate the initial privacy consumption more strictly than other privacy frameworks. This feature is particularly advantageous in privacy-sensitive applications where low privacy consumption in small iterations and failure probability are essential, such as in healthcare or financial analytics.

In instances with high iteration counts,  the performance of ADP requires careful evaluation due to the relatively large growth rate of privacy consumption. Simulation results show that the total privacy consumption under ADP can become significant as the number of iterations increases, especially when the failure probability $\delta$ is less restricted. Therefore, although ADP offers specific advantages in the initial stage, its overall privacy cost may exceed that of other differential privacy frameworks such as Rényi Differential Privacy (RDP) or zero-concentrated differential privacy (zCDP) during long-term iterations. Practitioners must carefully evaluate the iteration requirements and privacy constraints of their specific applications before choosing ADP as a privacy framework.

ADP offers a promising enhancement to conventional differential privacy models, providing refined privacy assurances that can be adjusted to satisfy particular needs. Nonetheless, its constraints in extensive iteration scenarios underscore the necessity of evaluating context-specific criteria while selecting among various privacy frameworks. Evaluating ADP's early advantages alongside its possible disadvantages over extended durations is a crucial factor in its effective application.

\subsection{Future Work}

Future research could focus on advancing the practical applications of ADP to enhance its robustness and adaptability in diverse privacy-preserving contexts. Expanding ADP beyond the Gaussian mechanism to include other mechanisms like the Laplace and Exponential mechanisms may provide insights into its flexibility across different data distributions and queries, reinforcing its role as a versatile privacy framework.

Evaluating ADP in practical settings such as healthcare, and finance will be crucial to determining its real-world utility and assessing how its theoretical benefits translate into practice. Understanding its performance amidst data heterogeneity, dynamic updates, and varying privacy requirements will be key to optimizing its deployment.

Moreover, integrating ADP into machine learning and deep learning systems could open up new possibilities for privacy-preserving models. This research could explore how ADP can be effectively incorporated into federated learning or privacy-preserving optimization while ensuring model accuracy and managing privacy consumption over multiple training iterations.

\end{document}